\documentstyle[12pt]{article}

\newcommand{\ket}{\rangle}
\newcommand{\bra}{\langle}
\begin{document}
\setlength{\baselineskip}{8mm}
\setlength{\jot}{3mm}
\begin{center}
{\Large\bf 
Moving Picture and Hamilton-Jacobi 
Theory in Quantum Mechanics}
\vspace{5cm}

Akihiro Ogura\footnote{ogu@mascat.nihon-u.ac.jp} 
and Motoo Sekiguchi${}^{\dagger}$\footnote{motoo@kokushikan.ac.jp}
 
\hspace{1cm}

Laboratory of Physics, Nihon University, Chiba 271-8587, Japan

and 

${}^{\dagger}$Faculty of Engineering, Kokushikan University, 
       Tokyo 154-8515, Japan

\vspace{1.5cm}

\begin{abstract}
We propose a new picture, which we call the {\it moving picture}, 
in quantum mechanics. The Schr\"{o}dinger equation in this picture 
is derived and its solution is examined. 
We also investigate the close relationship between the moving picture 
and the Hamilton-Jacobi theory in classical mechanics.

\end{abstract}
\end{center}
\newpage
\section{ Introduction }

Time development plays a fundamental role in quantum mechanics. 
In a general course on quantum mecanics, two 
well-known pictures are studied. One is the Schr\"{o}dinger picture, 
and the other is the Heisenberg picture. 
In the former, the operators are fixed in time and the states 
vary with time, while in the latter it is vice versa. 
In both pictures, we use only a stationary base. 

In contrast to this, we can also choose a 
set of base which acquires time dependency. 
The time-development of the base state is used 
in the path integral formulation of quantum mechanics. 
There, the Feynman 
propagator $K(x,t;x_{0},t_{0})=\bra x,t | x_{0},t_{0} \ket$ is the 
corner stone and the {\it moving frame} is defined as 
$|x,t \ket = e^{i \hat{H}t/\hbar}|x \ket$.  
In this paper, we will take this picture, which we call 
{\it moving picture}, and reformulate the quantum Hamilton 
formalism from this point of view. 

Recently, Omote et al.~\cite{omote} presented a new idea for the 
correspondence between classical and quantum mechanics. 
They proposed a new method for finding the solution to Schr\"{o}dinger 
equation from a classical canonical transformation for the case in 
which the transformed Hamiltonian becomes zero. 
Under this transformation, they fixed the tranformation of the 
canonical position $q \to Q(t)$ and the momentum $p \to P(t)$. 
Next, they made the tranformed position the operator $\hat{Q}(t)$ 
in the quantum mechanical sense, and the eigenstate $|Q;t \ket$ of 
the operator $\hat{Q}(t)$ with eigenvalue $Q$ form the set of base. 
In this representation, 
the Hamiltonian of the Schr\"{o}dinger equation becomes zero in the same 
way as for the Hamilton-Jacobi theory in classical mechanics. Thus, 
they called their formulation the {\it Hamilton-Jacobi picture} and 
also found a solution to the Schr\"{o}dinger equation. 
However, as we will show in this article, this 
{\it Hamilton-Jacobi picture} 
is nothing other than {\it moving picture}. In other words, their 
formulation is just a quantum mechanical formulation  with respect 
to the moving frame. 

Since the moving picture corresponds to looking at 
a moving body from a body-fixed moving reference frame, 
the transformed Hamiltonian always vanishes. This is a similar 
situation to the Hamilton-Jacobi theory in classical mechanics, 
where the canonical transformation makes the Hamiltonian become zero. 
The quantum Hamilton-Jacobi theory has been discussed 
by some authors~\cite{lee, kim}, but we will discuss more clearly 
the relationship between the moving picture in quantum mechanics and 
the Hamilton-Jacobi theory in classical mechanics. 

This article is organized as follows. In section 2, we formulate the 
{\it moving picture} and fix our notation. The Schr\"{o}dinger equation 
and its solution in this representation are derived in section 3, and 
some examples are discussed in section 4. In section 5, we will deduce 
the relationship between the moving picture and the Hamilton-Jacobi 
theory in classical mechanics. 
Section 6 is devoted to a discussion.

\section{ Moving Picture }

Time development in quantum mechanics 
is governed by the Hamiltonian $\hat{H}(t)$ for the system with 
the Schr\"{o}dinger equation for the time-evolution operator 
$\hat{T}(t)$, 
\begin{equation}
  i\hbar \frac{\partial}{\partial t} \hat{T}(t,t_{0}) 
  = \hat{H}(t) \hat{T}(t,t_{0}) .
\label{eq:timeevo}
\end{equation}
Here, $t_{0}$ is the initial time for the system and hereafter, 
for simplicity, $t_{0}=0$. $\hat{~}$ is the operator 
in quantum mechanics. 

When the solution to eq.(\ref{eq:timeevo}) is found, we define 
the unitary transformation of 
the position $\hat{q}$ and the momentum $\hat{p}$ operators as 
\begin{equation}
  \left\{
    \begin{array}{@{\,}ll}
        \hat{Q}(t) &=\hat{T}(t)\hat{q}\hat{T}^{\dagger}(t)  \\
        \hat{P}(t) &=\hat{T}(t)\hat{p}\hat{T}^{\dagger}(t) , 
    \end{array}
  \right.
\label{eq:transx}
\end{equation}
where the evolution of time is in the opposite direction to the 
Heisenberg operator. We note that since the right hand side is 
a function of position $\hat{q}$, momentum $\hat{p}$ and time $t$, 
the physical meaning of the left hand side, $\hat{Q}(t)$ 
and $\hat{P}(t)$, are not known at this point. 
It is easy to see that the transformed position and momentum operators 
have the commutation relation 
\begin{equation}
  \left[ \hat{Q}(t), \hat{P}(t) \right] = i \hbar ,
\label{eq:commu}
\end{equation}
if $\left[ \hat{q}, \hat{p} \right] = i \hbar$ is satisfied. 
This is closely related to classical mechanics, where 
the Poisson bracket is kept to 1 before and after the canonical 
transformation. 

Now we will make a complete set of base. 
We take the eigenstate $|Q;t \ket$ of the operator $\hat{Q}(t)$ with 
eigenvalue $Q$ to form the $Q$-representation:
\begin{equation}
  \hat{Q}(t)|Q;t \ket = Q|Q;t \ket. 
\label{eq:eigeneq}
\end{equation}
From the commutation relation eq.(\ref{eq:commu}), the following 
relationship 
\begin{equation}
  \bra Q;t|\hat{P}(t) 
              = -i\hbar \frac{\partial}{\partial Q} \bra Q;t|
\label{eq:prep}
\end{equation}
is satisfied. According to the transformation eq.(\ref{eq:transx}), 
the time evolution of the base is defined by 
\begin{equation}
  |Q;t \ket = \hat{T}(t)|Q \ket, \hspace{1cm} 
  \hat{q}|Q\ket=Q|Q\ket.
\label{eq:jika}
\end{equation}

The $q$-representation of the eigenstate $|Q;t \ket$ is calculated 
from eq.(\ref{eq:eigeneq}) 
\begin{equation}
  \bra q |\hat{Q}(t)|Q;t \ket = Q \bra q|Q;t \ket. 
\label{eq:diff}
\end{equation}
Since the transformed position operator $\hat{Q}(t)$ is a function 
of $\hat{q}$, $\hat{p}$ and $t$, this equation becomes 
the differential equation. With eq.(\ref{eq:prep}) and 
the normalized condition 
\begin{equation}
  \bra Q;t|Q';t \ket = \delta(Q-Q'), 
\end{equation}
we solve the differential equation (\ref{eq:diff}) and get 
the $q$-representation of the eigenstate $|Q;t \ket$ 
\begin{equation}
  \bra q|Q;t \ket. 
\label{eq:trans}
\end{equation}
This function will be used as the transformation function. 
We also note here that from eq.(\ref{eq:jika}) this function 
can be written by 
\begin{equation}
  \bra q|Q;t \ket = \bra q|\hat{T}(t)|Q \ket, 
\end{equation}
which is the Feynman propagator from the "position" $|Q\ket$ to 
the position $|q\ket$. This situation will be accomplished by 
some examples in the later section. 

\section{ Schr\"{o}dinger equation and its solution }

In the previous section, we have defined {\it moving picture}. 
Now we are ready to consider the Schr\"{o}dinger equation 
in this picture. 

\subsection{ Schr\"{o}dinger equation}

Let $|\Psi;t \ket_{\rm S}$ be the state which is a solution of 
the Schr\"{o}dinger equation
\begin{equation}
  i\hbar \frac{\partial}{\partial t}|\Psi;t \ket_{\rm S} 
               = \hat{H}(t)|\Psi;t \ket_{\rm S} ,
\end{equation}
where S stands for the Schr\"{o}dinger picture. 
We define the wave function in the moving picture as 
\begin{equation}
  \Psi(Q,t) = \bra Q;t|\Psi; t \ket_{\rm S} 
               = \bra Q|\hat{T}^{\dagger}(t)|\Psi; t \ket_{\rm S} .
\end{equation}
Then, the Schr\"{o}dinger equation becomes 
\begin{equation}
  i\hbar \frac{\partial}{\partial t} \Psi(Q,t) 
    = \int dQ' \bra Q|\hat{K}(t)|Q' \ket \Psi(Q',t) , 
\end{equation}
where the transformed Hamiltonian $\hat{K}(t)$ in the moving 
picture is 
\begin{equation}
  \hat{K}(t) = \hat{T}^{\dagger}(t) \hat{H}(t) \hat{T}(t) 
     + i\hbar \frac{\partial \hat{T}^{\dagger}(t)}{\partial t}\hat{T}(t). 
\end{equation}
But we notice that as with eq.(\ref{eq:timeevo}), the 
transformed Hamiltonian $\hat{K}(t)$ is identically zero: 
$\hat{K}(t)=0$. 
This is a similar situation to the Hamilton-Jacobi theory 
in classical mechanics. 
That is why Omote et al.~\cite{omote} have called this picture the 
{\it Hamilton-Jacobi picture}. However, this is nothing other than 
quantum mechanics with respect to the moving picture. 
It is certain that at first they seek a canonical transformation 
for which the Hamiltonian is zero in the region of classical mechanics, 
and then construct the representation for quantum mechanics. But it is 
unnecessary to go back to classical mechanics, as easily seen 
by the above argument.

\subsection{Arbitrary state in moving picture}

Since we have the transformation function eq.(\ref{eq:trans}), 
the wave function for an arbitrary state $|\psi \ket$ in the 
moving picture is easily obtained from 
\begin{equation}
  \bra Q;t|\psi \ket = \int dq \bra Q;t|q \ket \bra q|\psi \ket
\label{eq:arst}
\end{equation}
if the $q$-representation of an arbitrary state $\bra q|\psi \ket$ 
is known. 

It is worth commenting here that the state in the moving picture 
is independent of time. In fact, if the arbitrary state 
in the Schr\"{o}dinger picture is written by $|\psi;t\ket_{\rm S}$, 
its moving picture is 
\begin{equation}
  \bra Q;t|\psi;t\ket_{\rm S} 
            = \bra Q|\hat{T}^{\dagger}(t) \hat{T}(t)|\psi\ket_{\rm H}
            = \bra Q|\psi\ket_{\rm H} , 
\end{equation}
where H stands for the Heisenberg picture. 
This means that the {\it moving} representation for an arbitrary 
state in the Schr\"{o}dinger picture is equivalent to the 
$Q$-representation for 
an arbitrary state in the Heisenberg picture.

\section{ Examples }

We are now in a position to apply the moving picture 
to some systems. We take two examples and restrict ourselves to 
cases where the Hamiltonian does not depend on time. In this case, 
the Schr\"{o}dinger equation for the time-evolution operator 
eq.(\ref{eq:timeevo}) is easily calculated and we get
\begin{equation}
  \hat{T}(t) = \exp \left[ -\frac{i}{\hbar} \hat{H} t \right]. 
\end{equation}
However, the discussions in section 2 and 3 are applicable 
to all systems which satisfy the Schr\"{o}dinger equation 
for the time-evolution operator eq.(\ref{eq:timeevo}). 

\subsection{Free particle}

The Hamiltonian of a free particle is 
\begin{equation}
  \hat{H} = \frac{\hat{p}^{2}}{2m}, 
\end{equation}
where $m$ and $p$ are the mass and momentum of the particle. 
In this case, the time-evolution operator is 
\begin{equation}
  \hat{T}(t) = \exp \left[ -\frac{i}{\hbar}
               \frac{\hat{p}^{2}}{2m}t \right]. 
\end{equation}

It is easy to calculate the transformation of position and momentum, 
\begin{equation}
  \left\{
    \begin{array}{@{\,}ll}
        \hat{Q}(t) &=\hat{T}(t)\hat{q}\hat{T}^{\dagger}(t) 
                      = \hat{q} - \frac{t}{m} \hat{p} \\
        \hat{P}(t) &=\hat{T}(t)\hat{p}\hat{T}^{\dagger}(t) 
                      = \hat{p}.  
    \end{array}
  \right.
\label{eq:tra0}
\end{equation}
From a classical mechanical point of view, this canonical transformation 
is reproduced by the generating function 
\begin{equation}
  W(q, Q, t) = \frac{m}{2t}(q - Q)^{2} .
\label{eq:S0}
\end{equation} 
Furthermore, since this transformation is the Galilean 
transformation, the transformed Hamiltonain will vanish, 
as in the Hamilton-Jacobi theory. 
Also, since the operator $\hat{Q}$ is defined as a 
linear combination of $\hat{q}$, $\hat{p}$ and $t$, this transformation 
is neither a point transformation, nor a transformation from 
$q$ to $Q$, nor from $q$ to $p$.

Now we make up the set of base for the moving picture. From 
eq.(\ref{eq:diff}) and eq.(\ref{eq:tra0}), we obtain 
the differential equation 
\begin{equation}
  \bra q|\hat{Q}(t)|Q;t \ket 
   = \left[ q + i\hbar\frac{t}{m}\frac{\partial}{\partial q} \right]
   \bra q|Q;t \ket = Q \bra q|Q;t \ket .
\end{equation}
Integrating this equation, we have 
\begin{equation}
  \bra q|Q;t \ket = \Phi(Q) \sqrt{\frac{m}{2\pi i \hbar t}}
  \exp\left[ \frac{i}{\hbar}\frac{m}{t}
  \left( \frac{q^{2}}{2} - qQ \right) \right]
\end{equation}
where $\Phi(Q)$ is an arbitrary phase factor. On the other hand, 
the matrix element of an arbitrary state $|\psi\ket$ is 
\begin{eqnarray}
  \bra Q;t|\hat{P}|\psi \ket 
  &=& \int dq \bra Q;t|q\ket\bra q|\hat{P}|\psi\ket \\
  &=& \left[ -i\hbar\frac{\partial}{\partial Q} 
           + \frac{i\hbar}{\Phi^{\ast}(Q)}
                   \frac{\partial \Phi^{\ast}(Q)}{\partial Q}
            -\frac{m}{t}Q \right] \bra Q;t|\psi \ket.
\end{eqnarray}
If we take $\Phi^{\ast}(Q)=\exp\left[ -\frac{i}{\hbar}\frac{m}{t}
\frac{Q^{2}}{2} \right]$, the matrix element is given by 
\begin{equation}
  \bra Q;t|\hat{P}|\psi \ket
           = -i\hbar\frac{\partial}{\partial Q} \bra Q;t|\psi\ket
\end{equation}
and the transformation function turns out to be 
\begin{equation}
  \bra q|Q;t \ket = \bra q|\hat{T}(t)|Q \ket 
  = \sqrt{ \frac{m}{2\pi i \hbar t} } 
      \exp\left[ \frac{i}{\hbar} \frac{m}{2t}
        \left( q-Q \right)^{2} \right]. 
\label{eq:tqQ0}
\end{equation}
It is of interest that this transfromation function is nothing but 
the Feynman propagator from the "position" $|Q\ket$ to 
the position $|q \ket$.

As an example of an arbitrary state, we take a momentum eigenstate 
$|p \ket$ with eigenvalue $p$, which satisfies 
\begin{equation}
  \hat{p}|p \ket = p|p \ket.
\end{equation}
We calculate 
\begin{eqnarray}
  \bra Q;t|p \ket &=& \int dq \bra Q;t|q\ket\bra q|p\ket 
               = \bra Q|\hat{T}^{\dagger}(t)|p\ket \\ 
                  &=& \frac{1}{\sqrt{2\pi\hbar}} 
         \exp\left[ \frac{i}{\hbar}Q p 
         + \frac{i}{\hbar}\frac{p^{2}}{2m}t \right]. 
\end{eqnarray}

\subsection{Harmonic Oscillator}

The Hamiltonian of the harmonic oscillator is 
\begin{equation}
  \hat{H} = \frac{\hat{p}^{2}}{2m} +\frac{m \omega^{2}}{2}\hat{q}^{2}, 
\end{equation}
where $m$ and $\omega$ are the mass and frequency of the particle. 
Since this Hamiltonian is independent of time, the time-evolution 
operator is given by 
\begin{equation}
  \hat{T}(t) = \exp \left[ -\frac{i}{\hbar}
       \left\{\frac{\hat{p}^{2}}{2m} 
        + \frac{m \omega^{2}}{2}\hat{q}^{2} \right\}
         t \right]. 
\end{equation}

In the same manner, it is easy to calculate the transformed position 
and momentum, 
\begin{equation}
  \left\{
    \begin{array}{@{\,}ll}
        \hat{Q}(t) &=\hat{T}(t)\hat{q}\hat{T}^{\dagger}(t) 
                      = \hat{q} \cos \omega t 
                          - \frac{\hat{p}}{m \omega} \sin \omega t \\
        \hat{P}(t) &=\hat{T}(t)\hat{p}\hat{T}^{\dagger}(t) 
                      = m \omega \hat{q} \sin \omega t 
                         +\hat{p} \cos \omega t.  
    \end{array}
  \right.
\label{eq:traHO}
\end{equation}
From the classical mechanical point of view, this transformation 
is deduced from the generating function 
\begin{equation}
  W(q,Q,t) = \frac{m\omega}{\sin \omega t}
           \left(\frac{q^{2}+Q^{2}}{2} \cos \omega t-qQ  \right).
\label{eq:SHO}
\end{equation}
Furthermore, since this transformation refers to a rotational system in 
phase space, the transformed Hamiltonian will vanish, as with 
the Hamilton-Jacobi theory. 
Moreover, since the operator $\hat{Q}$ is defined as a 
linear combination of $\hat{q}$, $\hat{p}$ and $t$, this transformation 
is neither a point transformation, nor a transformation from 
$q$ to $Q$, nor from $q$ to $p$. 

Now we construct the set of base for the moving picture. From 
eq.(\ref{eq:diff}) and eq.(\ref{eq:traHO}), 
we obtain the differential equation
\begin{equation}
  \bra q|\hat{Q}(t)|Q;t \ket 
   = \left[ q\cos \omega t + \frac{i\hbar}{m\omega}\sin \omega t
      \frac{\partial}{\partial q} \right]\bra q|Q;t \ket 
    = Q \bra q|Q;t \ket .
\end{equation}
Integrating this equation, we have 
\begin{equation}
  \bra q|Q;t \ket 
      = \Phi(Q) \sqrt{\frac{m\omega}{2\pi i \hbar \sin \omega t}}
      \exp\left[ \frac{i}{\hbar}\frac{m\omega}{\sin \omega t}
  \left( \frac{q^{2}}{2}\cos\omega t - qQ \right) \right]
\end{equation}
where $\Phi(Q)$ is an arbitrary phase factor. On the other hand, 
the matrix element of an arbitrary state $\psi$ is 
\begin{eqnarray}
  \bra Q;t|\hat{P}|\psi \ket 
  &=& \int dq \bra Q;t|q\ket\bra q|\hat{P}|\psi \ket \\
  &=& \left[ -i\hbar\frac{\partial}{\partial Q} 
           + \frac{i\hbar}{\Phi^{\ast}(Q)}
                   \frac{\partial \Phi^{\ast}(Q)}{\partial Q}
            -m\omega Q\frac{\cos\omega t}{\sin\omega t} \right] 
        \bra Q;t|\psi \ket.
\end{eqnarray}
If we take $\Phi^{\ast}(Q)=\exp\left[ -\frac{i}{\hbar}
\frac{m\omega\cos\omega t}{2\sin\omega t}Q^{2} \right]$, 
the matrix element is given by 
\begin{equation}
  \bra Q ;t|\hat{P}|\psi \ket
           = -i\hbar\frac{\partial}{\partial Q} \bra Q;t|\psi\ket
\end{equation}
and the transformation function turns out to be 
\begin{equation}
  \bra q|Q;t \ket = \bra q|\hat{T}(t)|Q \ket 
  = \sqrt{\frac{m\omega}{2\pi i \hbar \sin\omega t}} 
      \exp\left[ \frac{i}{\hbar}\frac{m\omega}{\sin\omega t}
        \left( \frac{q^{2}+Q^{2}}{2}\cos\omega t-qQ \right) \right]. 
\label{eq:tqQHO}
\end{equation}
It is of interest that this transformation function is nothing other 
than the Feynman propagator from the "position" $|Q\ket$ to 
the position $|q \ket$.

As an example of an arbitrary state, we take a number eigenstate 
$|n \ket$, which satisfies 
\begin{equation}
  \hat{N}|n \ket = n|n \ket, 
\end{equation}
where the number operator is defined by 
$\hat{N} = \hat{a}^{\dagger} \hat{a}$. 
The operator $\hat{a}$ is related to the position and the momentum 
operators as $\hat{a}= \sqrt{\frac{m\omega}{2\hbar}}
\left( \hat{x}+\frac{i\hat{p}}{m\omega} \right)$.

In order to obtain the $Q$-representation $\bra Q;t|n\ket$, it is 
easier to proceed as follows rather than via eq.(\ref{eq:arst}). 
We first calculate 
\begin{eqnarray}
  \bra Q;t|\hat{a}^{\dagger}|\psi\ket 
  &=& \int dq \bra Q;t|q \ket\bra q|\hat{a}^{\dagger}|\psi\ket \\
  &=& \sqrt{\frac{m\omega}{2\hbar}} \int dq \bra Q;t|q \ket
      \left( q-\frac{\hbar}{m\omega}\frac{\partial}{\partial q} \right)
      \bra q|\psi\ket, 
\end{eqnarray}
which can be rewritten as 
\begin{equation}
    \bra Q;t|\hat{a}^{\dagger}|\psi\ket 
     = \sqrt{\frac{m\omega}{2\hbar}} e^{i\omega t} 
       \left(-\frac{\hbar}{m\omega}\right) 
       \exp\left[ \frac{m\omega}{2\hbar}Q^{2} \right] 
       \frac{\partial}{\partial Q} \left\{ -\exp\left[ 
      \frac{m\omega}{2\hbar}Q^{2} \right]\bra Q;t|\psi\ket \right\}.
\end{equation}
Further, a straight cast is done to get 
\begin{eqnarray}
  \bra Q;t|0\ket &=& \int\bra Q;t|q\ket\bra q|0\ket \\
     &=& \left( \frac{m\omega}{\pi\hbar} \right)^{1/4} 
     \exp\left[ -\frac{m\omega}{2\hbar}Q^{2}+i\frac{\omega}{2}t \right], 
\end{eqnarray}
thus we have 
\begin{eqnarray}
  \bra Q;t|n\ket &=& \frac{1}{\sqrt{n!}}
              \bra Q;t|\left(\hat{a}^{\dagger}\right)^{n}|0\ket \\
  &=& \left(\frac{m \omega}{\pi \hbar}\right)^{1/4} 
                       \frac{1}{\sqrt{2^{n} n!}} 
   \exp\left[ -\frac{m \omega}{2\hbar} Q^{2} 
              + i \left( n+\frac{1}{2} \right) \omega t \right]
    {\rm H}_{n}\left(\sqrt{\frac{m \omega}{\hbar}} Q \right), 
\end{eqnarray}
where H${}_{n}(\xi)$ is the Hermite polynomial of argument $\xi$.

The other example is a coherent state $\hat{a}|z\ket=z|z\ket$. 
The $Q$-representation is 
\begin{eqnarray}
  \bra Q;t|z\ket &=& \sum^{\infty}_{n=0} \bra Q;t|n\ket\bra n|z\ket \\
  &=& \left(\frac{m \omega}{\pi \hbar}\right)^{1/4} 
    \exp\left[ -\frac{m \omega}{2\hbar} Q^{2} 
              + 2zQe^{i\omega t}\sqrt{\frac{m\omega}{2\hbar}}
              -\frac{z^{2}}{2}e^{2i\omega t} - \frac{|z|^{2}}{2}
              +i\frac{\omega t}{2}
    \right] .
\end{eqnarray}

\section{ Principal function and moving picture }

Thus far, we have been discussing quantum mechanics from the 
perspective of the {\it moving} picture. In this case, since we are 
looking at 
the moving body from the body-fixed moving reference frame, 
the transformed Hamiltonian $\hat{K}(t)$ always vanishes. 
The situation is similar for the Hamilton-Jacobi theory 
in classical mechanics. 
In this section, we will study the moving picture from a different 
point of view. We start from the Schr\"{o}dinger equation, 
\begin{equation}
  i\hbar \frac{\partial}{\partial t} \psi(q,t) = 
   \left[ -\frac{\hbar^{2}}{2m}\frac{\partial^{2}}{\partial q^{2}}
          + V(q) \right] \psi(q,t) .
\label{eq:SE}
\end{equation}

Let us write the wave function as 
\begin{equation}
  \psi(q,t) = \exp\left[ \frac{i}{\hbar} S(q,t) \right], 
\label{eq:solu}
\end{equation}
and putting this into the above Schr\"{o}dinger equation, we get 
\begin{equation}
  \frac{1}{2m}\left(\frac{\partial S}{\partial q}\right)^{2} 
  + V(q) + \frac{\partial S}{\partial t} 
  -\frac{i\hbar}{2m}\frac{\partial^{2}S}{\partial q^{2}} = 0. 
\label{eq:HJ-0}
\end{equation}
This form of the equation has been well studied in the classical 
limit using the WKB formalism where  the focus is on the 
stationary-state solution. 
Here, we consider this equation from a different point of view. 
We define the new function~\cite{sakoda}
\begin{equation}
  F \equiv \frac{1}{2m}\frac{\partial^{2}S}{\partial q^{2}}. 
\label{eq:defF}
\end{equation}
If the function $S$ is given by the polynomial with respect to $q$ 
up to second order, $F$ is independent of $q$ and depends only on $t$. 
We consider the case hereafter. 
In this case, we define the new function $W$ as 
\begin{equation}
  S(q,t) = W(q,t) + i\hbar \int^{t} dt' F(t'). 
\label{eq:defS}
\end{equation}
Putting this back into eq.(\ref{eq:HJ-0}), we get 
\begin{equation}
  \frac{1}{2m}\left(\frac{\partial W}{\partial q}\right)^{2} 
  + V(q) + \frac{\partial W}{\partial t} = 0. 
\label{eq:HJ}
\end{equation}
This is the Hamilton-Jacobi equation that appears in classical mechanics. 

To sum up the above argument, once we have the solution $W$ of 
eq.(\ref{eq:HJ}),
we derive the function $F$ from eq.(\ref{eq:defF}) and also derive the 
function $S$ from eq.(\ref{eq:defS}). Accordingly, we get the solution 
to the Schr\"{o}dinger equation eq.(\ref{eq:SE}). 
Two examples will be discussed below. 

\subsection{Free particle}

Eq.(\ref{eq:S0}) is the solution of the Hamilton-Jacobi equation 
in classical mechanics, 
\begin{equation}
  \frac{1}{2m}\left(\frac{\partial W}{\partial q}\right)^{2} 
  + \frac{\partial W}{\partial t} = 0. 
\label{eq:HJ0}
\end{equation}
From eq.(\ref{eq:defF}), the function $F$ is 
\begin{equation}
  F = \frac{d}{dt} \ln \sqrt{t}, 
\end{equation}
then the solution to the Schr\"{o}dinger equation becomes 
\begin{equation}
  \psi = \exp\left[ \frac{i}{\hbar} S \right] 
       = \frac{1}{\sqrt{t}} \exp\left[ \frac{i}{\hbar} \frac{m}{2t} 
                     \left(q - Q \right)^{2} \right].
\end{equation}
This is the transformation function eq.(\ref{eq:tqQ0}) except for 
the arbitrary constant $\sqrt{\frac{m}{2\pi i \hbar }}$ which 
is calculated from normalization of the wave function. 

Next, we apply the Legendre transformation to eq.(\ref{eq:S0}), 
where the variables are transformed from $W(q, Q, t)$ 
to $W(q, P, t)$,  
\begin{eqnarray}
  W(q,P,t) &\equiv& W(q,Q,t) + Q P \\
                      &=& qP - \frac{P^{2}}{2m} t.
\label{eq:S0t}
\end{eqnarray}
This equation is also the solution to the Hamilton-Jacobi 
equation (\ref{eq:HJ0}). From eq.(\ref{eq:defF}), the function $F$ 
vanishes. Then the transformed solution (\ref{eq:solu}) becomes 
\begin{equation}
  \psi = \bra q|P;t\ket 
       = \frac{1}{\sqrt{2\pi\hbar}}\exp\left[ \frac{i}{\hbar} 
                      \left( qP-\frac{P^{2}}{2m}t \right) \right] .
\label{eq:tqQ0t}
\end{equation}
The canonical transformation in quantum mechanics was studied in the 
early days of quantum mechanics~\cite{dirac}, and has 
recently been reconsidered by some authors~\cite{lee, kim}. 
As already pointed out in~\cite{kim}, 
it is interesting that the generating functions
eq.(\ref{eq:S0}) and eq.(\ref{eq:S0t}) are transformed by 
the Legendre transformation, while the wave functions 
eq.(\ref{eq:tqQ0}) and eq.(\ref{eq:tqQ0t}) are transformed 
by the Fourier transformation, which is easily accomplished by 
\begin{eqnarray}
  \bra q|P;t\ket &=& \int dQ \bra q|Q;t\ket\bra Q;t|P;t\ket \\
                &=& \int dQ \bra q|Q;t\ket \times 
                \frac{1}{\sqrt{2\pi \hbar}}e^{iPQ/\hbar}.
\end{eqnarray}

\subsection{Harmonic Oscillator}

Eq.(\ref{eq:SHO}) is the solution of the Hamilton-Jacobi equation 
in classical mechanics, 
\begin{equation}
  \frac{1}{2m}\left(\frac{\partial W}{\partial q}\right)^{2} 
  + \frac{m \omega^{2}}{2} q^{2}
  + \frac{\partial W}{\partial t} = 0. 
\end{equation}

In this case, the function $F$ becomes 
\begin{equation}
  F = \frac{d}{dt} \ln \sqrt{\sin \omega t}, 
\end{equation}
and then the solution to the Schr\"{o}dinger equation is 
\begin{equation}
  \psi = \frac{1}{\sqrt{\sin \omega t}} 
       \exp\left[  \frac{i}{\hbar}
           \frac{m \omega}{\sin \omega t} 
          \left( \frac{q^{2}+Q^{2}}{2} \cos \omega t -qQ \right) \right]. 
\end{equation}
This is the transformation function eq.(\ref{eq:tqQHO}) except for 
the arbitrary constant $\sqrt{\frac{m\omega}{2\pi i \hbar }}$ which 
is calculated from normalization of the wave function.

In a similar manner, we apply the Legendre transformation: 
$W(q,Q,t) \to W(q,P,t)$, and obtain 
\begin{eqnarray}
  W(q,P,t) &\equiv& W(q,Q,t) + Q P \\
         &=& \frac{qP}{\cos \omega t}
              -\left( \frac{m\omega^{2}}{2}q^{2}
                    +\frac{P^{2}}{2m} \right) 
                   \frac{\tan \omega t}{\omega} .
\label{eq:SHOt}
\end{eqnarray}
From this equation, the function $F$ becomes
\begin{equation}
  F = \frac{d}{dt} \ln \sqrt{\cos \omega t}
\end{equation}
then the solution to the transformed Schr\"{o}dinger equation becomes
\begin{equation}
  \psi =\bra q|P;t\ket 
= \frac{1}{\sqrt{2\pi\hbar\cos \omega t}} 
         \exp\left[ \frac{i}{\hbar} 
                    \left\{ \frac{qP}{\cos \omega t}-
                    \left( \frac{m\omega^{2}}{2}q^{2}
                    +\frac{P^{2}}{2m} \right)\frac{\tan \omega t}{\omega}
               \right\} \right].
\label{eq:tqQHOt}
\end{equation}

The same argument for the transformation is also applied to 
the Harmonic Oscillator case. That is, 
the generating functions
eq.(\ref{eq:SHO}) and eq.(\ref{eq:SHOt}) are transformed by 
the Legendre transformation, while the wave functions 
eq.(\ref{eq:tqQHO}) and eq.(\ref{eq:tqQHOt}) are transformed 
by the Fourier transformation.

\section{ Discussion }

We have investigated the {\it moving picture} in quantum mechanics 
and can clearly formulate a representation of this picture. 

In contrast to the Heisenberg and Schr\"{o}dinger pictures, we found 
that the transformed Hamiltonian becomes zero. This is similar to 
the case of the Hamilton-Jacobi theory in classical mechanics. 
It is unsuitable to use the {\it Hamilton-Jacobi representation} 
terminology for this picture because the Hamilton-Jacobi theory 
permits a wider variety of transformations than in 
the quantum case. For example, we take a harmonic oscillator. 
The following transformation 
\begin{equation}
  \left\{
    \begin{array}{@{\,}ll}
        Q & = \frac{1}{\omega} \tan^{-1}
                       \left(\frac{m\omega q}{p}\right)-t \\
        P & = \frac{p^{2}}{2m}+\frac{m \omega^{2}}{2} q^{2}
    \end{array}
  \right.
\end{equation}
is really a canonical transformation and its transformed Hamiltonian 
becomes zero. But in this case, we never formulate a quantum 
representation 
for this transformation, because the operators $\hat{q}$ and $\hat{p}$ 
are included in an arctangent function. 

We also discussed the relationship between the {\it moving} picture 
and the Hamilton-Jacobi theory in classical mechanics. This shows 
a new correspondence between classical and quantum mechanics.

\vspace{1cm}

{\bf\large Acknowledgment}

We thank Professor Kamefuchi for useful discussions. 


\end{document}